\documentclass[seceq]{ptptex}

\usepackage{graphicx}
\usepackage{epsfig}

\title{%        %You can use \\ for explicit line-break
On the Discrepancy of $pp$, $\bar pp$ Total Cross Sections
at $\sqrt s = 1.8$TeV between E710, E811 and CDF
}

\author{%       %Use \scshape  for the family name
Keiji \textsc{Igi} and Muneyuki \textsc{Ishida}$^a$ 
}

\inst{%         %Affiliation, neglected when [addenda] or [errata]
Theoretical Physics Laboratory, RIKEN, Wako, Saitama 351-0198, Japan\\
$^a$Department of Physics, School of Science and Engineering, Meisei University, 
Hino, Tokyo 191-8506, Japan
}

\abst{%         %this abstract is neglected when [addenda] or [errata]
Based on the previous approach, we have investigated a possibility to
resolve the discrepancy between the E710, E811 and CDF at $\sqrt s =1.8$TeV,
using the experimental data of the $pp$, $\bar pp$ total cross sections 
$\sigma_{\rm tot}^{(+)}$ and $\rho^{(+)}$ ratio up to the SPS 
experiments ($\sqrt s = 0.9$TeV) as inputs.
We predict $\sigma_{\rm tot}^{\bar pp}$ and $\rho^{\bar pp}$ at the Tevatron energy
($\sqrt s=1.8$TeV) as  $\sigma_{\rm tot}^{\bar pp}=75.9\pm 1.0$mb, $\rho^{\bar pp}=0.136\pm 0.005$. 

    It turns out that only  the data of E710 is consistent with the
prediction in the one standard deviation. 
So we can conclude that E710 is preferable but we can exclude neither CDF nor E811 results.
}

\begin{document}

\maketitle

\section{Introduction}

Recently\cite{[1]}, we have searched for the simultaneous best fit of the
average of $\bar pp$, $pp$ total cross sections( $\sigma_{\rm tot}^{(+)}$ ), and the ratio
of the real to imaginary part of the forward scattering amplitude( $\rho^{(+)}$ )
for 70GeV $<$ $P_{lab}$ $<$ $P_{large}$ as inputs in terms  of high-energy
parameters $c_0$, $c_1$, $c_2$ and $\beta_{P^\prime}$ constrained by the FESR  
with $N(\simeq 10$GeV).
Block and Halzen\cite{[2],BH} also reached to the similar conclusions independently
based on duality in a different approach. We first chose $P_{large} =2100$GeV
corresponding to the ISR region( $\sqrt s \simeq 60$GeV ). Secondly we chose 
$P_{large}=2\times 10^6$GeV corresponding to the Tevatron energy( $\sqrt s \simeq 2$TeV ). 
We then predicted $\sigma_{\rm tot}^{(+)}$ and $\rho^{(+)}$ at the LHC and the high-energy
cosmic-ray energy regions.  It turned out that the prediction of $\sigma_{\rm tot}^{(+)}$ 
agrees with $pp$ experimental data at the cosmic-ray regions\cite{[3],[4],[5]} within
errors in the first case( ISR ). It has to be noted that the energy range of
predicted $\sigma_{\rm tot}^{(+)}$, $\rho^{(+)}$ is several orders of magnitude larger
than the energy regions of $\sigma_{\rm tot}^{(+)}$, $\rho^{(+)}$ input. 
If we use data up to Tevatron( the second case ), the situation has been much improved
although there are some systematic uncertainty coming from discrepancy of
the data between E710\cite{E710}, E811\cite{E811} and CDF\cite{CDF} at $\sqrt s =1.8$TeV\cite{[1]}. 
Finally we concluded that the precise measurements of $\sigma_{\rm tot}^{pp}$  
in the coming LHC experiments will resolve this discrepancy at $\sqrt s = 1.8$TeV.

The purpose of this paper is to investigate a possibility to resolve
this discrepancy using the experimental data of $\sigma_{\rm tot}^{(+)}$ and $\rho^{(+)}$
up to the SPS experiments ($\sqrt s = 0.9$TeV).

\section{The general approach}

As in the previous paper\cite{[1]}, let us first consider
the crossing-even forward scattering amplitude defined by
\begin{eqnarray}
F^{(+)}(\nu ) &=& \frac{f^{\bar pp}(\nu )+f^{pp}(\nu )}{2}\ \    
{\rm with}\ \  Im\ F^{(+)}(\nu )=\frac{k\ \sigma^{(+)}_{\rm tot}(\nu )}{4\pi}\ .
\label{eq1}
\end{eqnarray}

We also assume 
\begin{eqnarray}
Im\ F^{(+)}(\nu ) &=& Im\ R(\nu )+ Im\ F_{P^\prime}(\nu )\nonumber\\
 &=& \frac{\nu}{M^2}\left( c_0 + c_1 {\rm log}\frac{\nu }{M} + c_2 {\rm log}^2\frac{\nu }{M}  \right)
  + \frac{\beta_{P^\prime}}{M}\left( \frac{\nu}{M} \right)^{\alpha_{P^\prime}}\ \ \ \ \ 
\label{eq2}
\end{eqnarray}
at high energies ($\nu > N$).  
It is to be noted that $c_0$, $c_1$, $c_2$ and $\beta_{P^\prime}$ are dimensionless.
We have defined the functions $R(\nu )$ and $F_{P^\prime} (\nu )$ 
by replacing $\mu$ by M in Eq.~(3) of ref.~\citen{[6]}.
Here, $M$ is the proton( anti-proton) mass and $\nu ,\ k$ are the incident proton(anti-proton) 
energy, momentum in the laboratory system, respectively.

Since the amplitude is crossing-even, we have
\begin{eqnarray}
R(\nu ) &=& \frac{i\nu}{2M^2}\left\{ 2c_0+c_2\pi^2 
  + c_1 \left({\rm log}\frac{e^{-i\pi}\nu }{M}+{\rm log}\frac{\nu}{M}\right) \right. \nonumber\\
&& \left.  + c_2 \left({\rm log}^2\frac{e^{-i\pi}\nu }{M} + {\rm log}^2\frac{\nu}{M}\right)  \right\}\ ,\ \ \ \ \ \ \ 
 \label{eq3}\\
F_{P^\prime}(\nu ) &=& -\frac{\beta_{P^\prime}}{M}
 \left( \frac{(e^{-i\pi}\nu /M)^{\alpha_{P^\prime}}
       +(\nu /M)^{\alpha_{P^\prime}}}{{\rm sin}\pi\alpha_{P^\prime}} \right),
\label{eq4}
\end{eqnarray}
and subsequently obtain 
\begin{eqnarray}
Re\ R(\nu ) &=& \frac{\pi\nu}{2M^2}\left( 
  c_1 + 2 c_2 {\rm log}\frac{\nu}{M} \right)\ ,\ \ \  
 \label{eq5}\\
Re\ F_{P^\prime}(\nu ) &=& -\frac{\beta_{P^\prime}}{M}
 \left( \frac{\nu}{M}\right)^{0.5}\ ,\ \ \ 
\label{eq6}
\end{eqnarray}
substituting  $\alpha_{P^\prime} =\frac{1}{2}$ in Eq.~(\ref{eq4}).\\

\underline{FESR}: The FESR corresponding to $n=1$ \cite{[7],[8]} is:
\begin{eqnarray}  &&
\int_0^M \nu Im\ F^{(+)}(\nu )d\nu 
     + \frac{1}{4\pi}\int_0^{\overline{N}} k^2 \sigma_{\rm tot}^{(+)}(k)dk \nonumber\\
 & &=  \int_0^N \nu Im\ R(\nu ) d\nu 
     + \int_0^N \nu Im\ F_{P^\prime}(\nu ) d\nu \ \ \ . \ \ \  
\label{eq7}
\end{eqnarray}
We call Eq.~(\ref{eq7}) as the FESR which we use in our analysis.\\

\underline{The  $\rho^{(+)}$ ratio}: The $\rho^{(+)}$ ratio, 
the ratio of the real to imaginary part of 
 $F^{(+)}(\nu )$ was obtained from Eqs.~(\ref{eq2}), (\ref{eq5}) and (\ref{eq6}) as
\begin{eqnarray}
\rho^{(+)}(\nu ) &=& \frac{Re\ F^{(+)}(\nu )}{Im\ F^{(+)}(\nu )}
  = \frac{Re\ R(\nu )+Re\ F_{P^\prime}(\nu )}{Im\ R(\nu )+Im\ F_{P^\prime}(\nu )} \nonumber\\
  &=& \frac{ \frac{\pi\nu}{2M^2}\left( c_1+2c_2 {\rm log} \frac{\nu}{M} \right) 
          -\frac{\beta_{P^\prime}}{M}\left(\frac{\nu}{M}\right)^{0.5} }{
                  \frac{k\sigma_{\rm tot}^{(+)}(\nu)}{4\pi} }\ .\ \ \ 
\label{eq8}
\end{eqnarray}
Although the numerator of Eq.~(\ref{eq8}) becomes large for large values of $\nu$,
a real constant has to be introduced in principle since the dispersion
relation for $Re~F^{(+)}(\nu )$
requires a single subtraction constant $F^{(+)}(0)$\cite{[9],[2]}. So,
we also add $F^{(+)}(0)$ in the numerator as
\begin{eqnarray}
\rho^{(+)}(\nu ) &=&     \frac{ \frac{\pi\nu}{2M^2}\left( c_1+2c_2 {\rm log} \frac{\nu}{M} \right) 
          -\frac{\beta_{P^\prime}}{M}\left(\frac{\nu}{M}\right)^{0.5} + F^{(+)}(0) }{
                  \frac{k\sigma_{\rm tot}^{(+)}(\nu)}{4\pi} }\ .\ \ \ 
\label{eq9}
\end{eqnarray}
As will be discussed in the Appendix, the introduction of this constant
slightly
modifies the value of $\rho^{(+)}(\nu )$ although it will not affect the value of
$\sigma_{\rm tot}^{(+)}$. So, we use the Eq.~(\ref{eq9}) as the value of 
$\rho^{(+)}(\nu )$ in this analysis.

The FESR, Eq.~(\ref{eq7}), has some problem. i.e., there are the so-called 
unphysical regions coming from boson poles below the $\bar pp$ threshold.
So, the contributions from unphysical regions of the first term of the right-hand side
of Eq.~(\ref{eq7}) have to be calculated.
These contributions can be estimated to be an order of
0.1\% compared 
with the second term.\cite{[1]}
Thus, it can easily be neglected.

Therefore, the FESR, 
 the formula of $\sigma_{\rm tot}^{(+)}$(Eqs.~(\ref{eq1}) and (\ref{eq2})) 
 and the $\rho^{(+)}$ ratio (Eq.~(\ref{eq9})) are our starting points.
Armed with the FESR, we express high-energy parameters
$c_0,c_1,c_2,\beta_{P^\prime}$ in terms of the integral of total cross sections up to
$N$. 
Using this FESR as a constraint for $\beta_{P^\prime}=\beta_{P^\prime}(c_0,c_1,c_2)$,
there are four independent parameters including $F^{(+)}(0)$.
We then search for the simultaneous best fit to the data points of $\sigma_{\rm tot}^{(+)}(k)$
and $\rho^{(+)}(k)$ for 70GeV$\le k \le P_{large}$ corresponding to the SPS energy 
($P_{large}\simeq 0.43\times 10^6$GeV ($\sqrt s=0.9$TeV)),
to determine the values of $c_0,c_1,c_2$ and $F^{(+)}(0)$ giving the least $\chi^2$. 
We thus predict the $\sigma_{\rm tot}$ and $\rho^{(+)}$ 
in the Tevatron energy region ($\sqrt s=1.8$TeV).

\section{Predictions for $\sigma_{\rm tot}^{(+)}$ and $\rho^{(+)}$ at $\sqrt s =
1.8$TeV}

    Using the data up to $\sqrt s = 0.9$TeV ( SPS ), we predict $\sigma_{\rm tot}^{(+)}$
and $\rho^{(+)}$ at the Tevatron energy ( $\sqrt s = 1.8$TeV ).\\

\underline{Analysis 1}:\ \ \  As was explained in the general approach (\S 2), 
both $\sigma_{\rm tot}^{(+)}$ and $Re~F^{(+)}$ data
in  $70{\rm GeV} < k < P_{large} = 4.3\times 10^5{\rm GeV} (\sqrt s = 0.9{\rm TeV}$ ) 
are fitted simultaneously through the formula of 
$\sigma_{\rm tot}^{(+)}$ ( Eqs.~(\ref{eq1}) and (\ref{eq2}))
and the $\rho^{(+)}$ ratio ( Eq.~(\ref{eq9})) with the FESR ( Eq.~(\ref{eq7})) as a
constraint.

The $\sigma_{\rm tot}^{(+)}(k)$ data points are obtained by averaging 
$\sigma^{\bar pp}_{\rm tot}$ and $\sigma^{pp}_{\rm tot}$ data points\cite{PDG} 
when they are listed at the same value of $k$. For the details of data treatment 
of $\sigma_{\rm tot}^{(+)}$ and $Re~F^{(+)}$, see ref.~\citen{[1]}. 
The FESR gives us 
\begin{eqnarray}
8.87 &=& c_0 + 2.04 c_1 + 4.26 c_2 + 0.367 \beta_{P^\prime}
\label{eqFESR}
\end{eqnarray}
(Eq.~(12) of ref.~\citen{[1]}),
where we use the central value of $\frac{1}{4\pi}\int_0^{\overline{N}} 
k^2 \sigma_{\rm tot}^{(+)}(k)=3403\pm 20$GeV\footnote{
This value is obtained by numerically 
integrating the experimental $k^2 \sigma_{\rm tot}^{(+)}=k^2 (\sigma_{\rm tot}^{\bar pp}
+\sigma_{\rm tot}^{pp})/2$. See, ref.~\citen{[1]} for details.
} 
for $\overline{N}=10$GeV in Eq.~(\ref{eq7}). 

\begin{figure}
\includegraphics{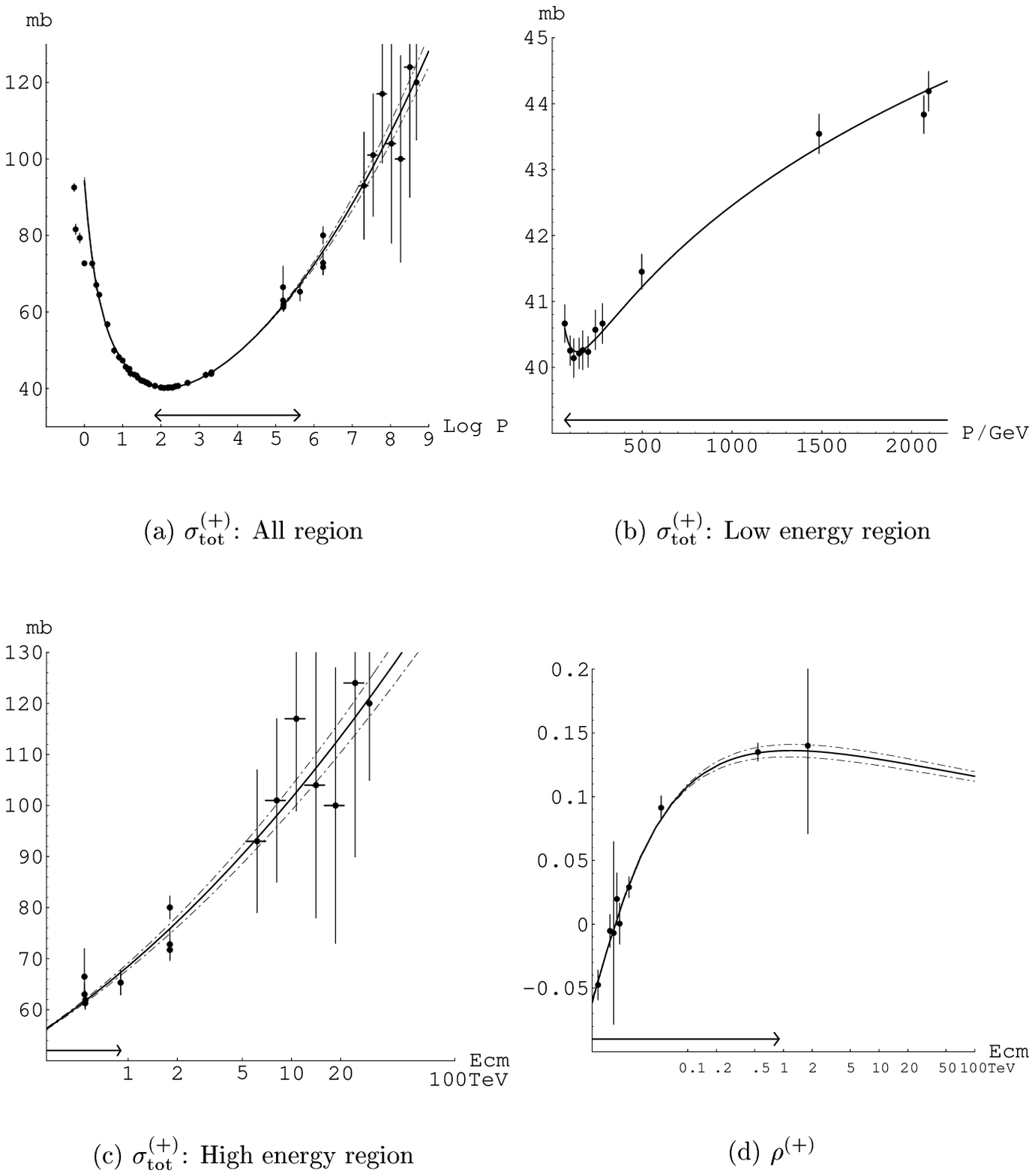}% Here is how to import EPS art
\caption{\label{fig1}
Predictions for $\sigma^{(+)}$ and $\rho^{(+)}$ in terms of 
the Analysis 1.
The fit is done for the data up to SPS energy, in the region  
70GeV$\le$ $k$ $\le$ $4.3\times10^5$GeV(11.5GeV $\le \sqrt s \le$ 0.9TeV) 
which is shown by the arrow. 
Total cross section $\sigma^{(+)}_{\rm tot}$ in 
(a) all energy region, versus log$_{10}P_{lab}/$GeV,
(b) low energy region (up to ISR energy), versus $P_{lab}/$GeV and
(c) high energy (Tevatron-collider, LHC and cosmic-ray energy) region, 
    versus center of mass energy $E_{cm}$ in TeV unit.
(d) gives the $\rho^{(+)}(=Re\ F^{(+)}/Im\ F^{(+)})$ in high energy region, 
    versus $E_{cm}$ in terms of TeV. 
The thin dot-dashed lines represent the one standard deviation
of $c_2$.(See the caption in Table \ref{tab1}.) The corresponding values of parameters are
$(c_2,c_1,c_0,\beta_{P^\prime},F^{(+)}(0))=
(0.0466\pm 0.0047,-0.161\mp 0.077,6.27\pm 0.31,7.45\mp 0.48,12.65\pm 0.69)$.
 }
\end{figure}

The result of the fit is shown in Fig.~\ref{fig1}.
The values of parameters and resulting $\chi^2$ are given in Tables \ref{tab1} and \ref{tab2},
respectively. 

\begin{table}
\caption{
The values of parameters in the best fit to the data up to SPS energy ($\sqrt s=0.9$TeV) 
in the analysis 1(fit to the data in $70{\rm GeV} < k < P_{large} = 4.3\times 10^5{\rm GeV}$ ).   
The error estimations are done as follows: The $c_2$ is fixed with 
a value deviated a little from the best-fit value, and then the $\chi^2$-fit 
is done by three parameters $c_0$, $c_1$ and $F^{(+)}(0)$,
where $\beta_{P^\prime}$ is represented by the other parameters through FESR(Eq.~(\ref{eqFESR})). 
When the resulting $\chi^2$ is larger than the least $\chi^2$ of the four-parameter fit by one, 
the corresponding value of $c_2$ gives one standard deviation. 
The higher and lower dot-dashed lines in Fig.~\ref{fig1} 
represent this deviation of $c_2$.
The errors of the other parameters are estimated through similar procedures.
}
\begin{center}
\begin{tabular}{c|ccccc|}
       & $c_2$  & $c_1$  & $c_0$  &  $\beta_{P^\prime}$ & $F^{(+)}(0)$ \\
\hline
Analysis 1 & $0.0466\pm 0.0047$ & $-0.161\mp 0.078$ & $6.27\pm 0.33$ & $7.45\mp 0.51$ & $12.65\pm 5.66$\\
\hline
\end{tabular}
\end{center}
\label{tab1}
\end{table}

\begin{table}
\caption{ The values of $\chi^2$ for the fit to data in 
$70{\rm GeV} < k < P_{large} = 4.3\times 10^5{\rm GeV}$(Analysis 1): 
$N_F$ and $N_\sigma (N_\rho )$ are the degree of freedom and 
the number of $\sigma^{(+)}_{\rm tot}(\rho^{(+)})$ data points in the fitted energy region.
}
\begin{center}
\begin{tabular}{c|ccc|}
         & $\chi^2/N_F$        
            &  $\chi^2_{\sigma}/N_{\sigma}$ & $\chi^2_{\rho}/N_{\rho}$   \\
\hline
Analysis 1  &  8.1/20 & 5.7/17 & 2.4/8\\
\hline
\end{tabular}
\end{center}
\label{tab2}
\end{table}

In terms of the best-fit values of parameters in Table \ref{tab1}
the predictions at $\sqrt s = 1.8$TeV are obtained as
\begin{eqnarray}
\sigma_{\rm tot}^{(+)} &=&  75.9 \pm 1.0\ {\rm mb},\ \   
\rho^{(+)} = 0.136 \pm 0.005\ , \ \
\label{eq10}
\end{eqnarray}
where the errors correspond to the 
one standard deviation of $c_2$, since
the $c_2$log$^2(\nu /M)$-term in Eq.~(\ref{eq2}) is most relevant for predicting 
$\sigma_{\rm tot}^{(+)}$ in high energy region. (See the caption in Table \ref{tab1}.)

The equation (\ref{eq10}) has to be compared with the experimental values at $\sqrt s =
1.8$TeV;
\begin{eqnarray}
\sigma_{\rm tot}^{\bar pp}(E811)  &=&  71.71 \pm 2.02\ {\rm mb},\ \nonumber\\   
\sigma_{\rm tot}^{\bar pp}(E710)  &=&  72.8 \pm 3.1\ {\rm mb},\ \nonumber\\   
\sigma_{\rm tot}^{\bar pp}(CDF)  &=&  80.03 \pm 2.24\ {\rm mb},\ \  
\label{eq11}
\end{eqnarray}
where we note that the difference between $\sigma_{\rm tot}^{\bar pp}$ and 
$\sigma_{\rm tot}^{(+)}$ is negligible at the relevant energy. 
It is worthwhile to notice that only the data of E710\cite{E710} is consistent with the
prediction, Eq.~(\ref{eq10}) in the one standard deviation ( $72.8+3.1=75.9$ ).

    If one tolerates two standard deviations, both CDF\cite{CDF} ($80.03-2.24\times 2=75.55$)
and E811\cite{E811}($71.71+2.02\times 2=75.75$) are consistent with the predictions Eq.~(\ref{eq10}).  
So we can conclude that E710 is preferable but we can exclude neither CDF nor
E811 results.

    The predictions at LHC energy ($\sqrt s = 14$TeV) in terms of the best fit
values of high-energy parameters in Table \ref{tab1} are
\begin{eqnarray}
\sigma_{\rm tot}^{pp} &=&  107.2 \pm 2.8\ {\rm mb},\ \   
\rho^{pp} = 0.128 \pm 0.005\ , \ \
\label{eq12}
\end{eqnarray}
where the errors correspond to one standard deviation of $c_2$.
We should note that Eq.~(\ref{eq12}) is consistent with the recent prediction by 
Block and Halzen\cite{BH},
$\sigma_{\rm tot}^{pp} =  107.3 \pm 1.2\ {\rm mb},\ \   
\rho^{pp} = 0.132 \pm 0.001$.\\

\underline{An interesting observation}:\ \ \ \ 
    We can make the following interesting observation.
We fitted
the data for $\sigma_{\rm tot}^{(+)}$ and $\rho^{(+)}$ above 70GeV, 
as is shown by the arrow
in the Fig.~1(a), Fig.~1(d) to predict higher-energy data. 
It is interesting
to observe that the prediction of $\sigma_{\rm tot}^{(+)}$ are also in good agreement
with experiments, even below 70GeV. 
The reason is as follows: The requirement of FESR,
Eq.~(\ref{eq7}) is nearly equal to require that the theoretical value of $\sigma_{\rm tot}^{(+)}$ is
nearly equal to the experimental value at the upper limit of the integral
$N=10$GeV since higher side of the integral is enhanced because of $k^2$
in the integral.

Because of this observation, we can apply the same formula to fit the data in the lower energy region
than in the analysis 1.\\

\underline{Analysis 2}:\ \ \ \ Data
in  $10{\rm GeV} < k < P_{large} = 4.3\times 10^5{\rm GeV} 
     (4.54{\rm GeV} < \sqrt s < 0.9{\rm TeV}$ ) 
are fitted through the same formula in the analysis 1.
Additionally 15(2) data points are included in $\sigma_{\rm tot}^{(+)}\ ( Re~F^{(+)} )$. 

\begin{figure}
\includegraphics{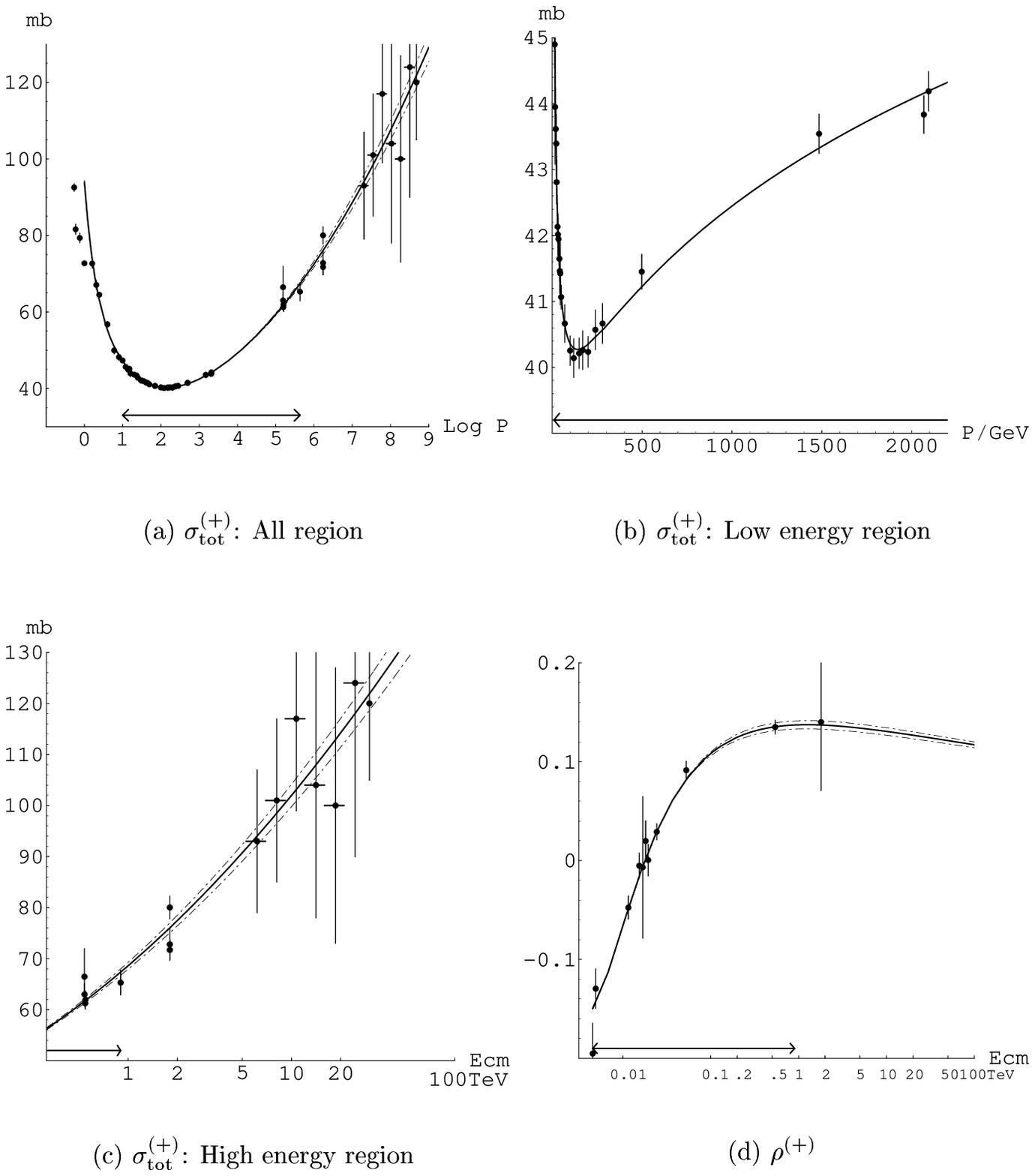}% Here is how to import EPS art
\caption{\label{fig2}
Predictions for $\sigma^{(+)}$ and $\rho^{(+)}$ in terms of 
the Analysis 2.
The fit is done for the data up to SPS energy, in the region  
10GeV$\le$ $k$ $\le$ $4.3\times10^5$GeV(4.54GeV $\le \sqrt s \le$ 0.9TeV) 
which is shown by the arrow. 
For each figure, see the caption in Fig.~\ref{fig1}.
The thin dot-dashed lines represent the one standard deviation
of $c_2$.(See the caption in Table \ref{tab1}.) The corresponding values of parameters are
$(c_2,c_1,c_0,\beta_{P^\prime},F^{(+)}(0))=
(0.0479\pm 0.0037,-0.186\mp 0.056,6.38\pm 0.21,7.26\mp 0.31,10.19\pm 0.31)$.
  }
\end{figure}

The result of the fit is shown in Fig.~\ref{fig2}.
The values of parameters and resulting $\chi^2$ are given in Tables \ref{tab3} and \ref{tab4},
respectively. 

\begin{table}
\caption{
The values of parameters in the best fit to the data up to the SPS energy ($\sqrt s=0.9$TeV) 
in the analysis 2(fit to the data in $10{\rm GeV} < k < P_{large} = 4.3\times 10^5{\rm GeV}$ ).   
We obtain smaller error of $F^{(+)}(0)$
than in analysis 1(Table \ref{tab1}), since, as is seen in Eq.~(\ref{eq9}), 
$F^{(+)}(0)$ has sizable effects only in the low energy region. 
For errors, see the caption in Table \ref{tab1}.
}
\begin{center}
\begin{tabular}{c|ccccc|}
       & $c_2$  & $c_1$  & $c_0$  &  $\beta_{P^\prime}$ & $F^{(+)}(0)$ \\
\hline
Analysis 2 & $0.0479\pm 0.0037$ & $-0.186\mp 0.057$ & $6.38\pm 0.22$ & $7.26\mp 0.33$ & $10.19\pm 1.72$\\
\hline
\end{tabular}
\end{center}
\label{tab3}
\end{table}

\begin{table}
\caption{ The values of $\chi^2$ for the fit to data in 
$10{\rm GeV} < k < P_{large} = 4.3\times 10^5{\rm GeV}$(Analysis 2).
For  
$N_F$ and $N_\sigma (N_\rho )$, see the caption in Table.~\ref{tab2}. 
}
\begin{center}
\begin{tabular}{c|ccc|}
         & $\chi^2/N_F$        
            &  $\chi^2_{\sigma}/N_{\sigma}$ & $\chi^2_{\rho}/N_{\rho}$   \\
\hline
Analysis 2  &  14.1/37 & 8.8/32 & 5.3/10\\
\hline
\end{tabular}
\end{center}
\label{tab4}
\end{table}

    The predictions at LHC energy ($\sqrt s = 14$TeV) in terms of the best fit
values of high-energy parameters in Table \ref{tab3} are
\begin{eqnarray}
\sigma_{\rm tot}^{pp} &=&  107.8 \pm 2.4\ {\rm mb},\ \   
\rho^{pp} = 0.129 \pm 0.004\ , \ \
\label{eq13}
\end{eqnarray}
where the errors correspond to the one standard deviation of $c_2$.
Essentially the same prediction are obtained as Eq.~(\ref{eq12}) of the analysis 1,
although the errors are slightly smaller.  
Our result is stable independently of the choices of the fitting energy range.

\section{Concluding remarks}

    In  \S 3, we have investigated a possibility to resolve the discrepancy
between E710, E811 and CDF, using the experimental data of $\sigma_{\rm tot}^{(+)}$
and $\rho^{(+)}$ up to the SPS experiments ($\sqrt s = 0.9$TeV).

    We came to the conclusion that only the data of E710 is consistent with
the prediction, Eq.~(\ref{eq10}) in the one standard deviation although we can 
exclude neither CDF nor E811 results in the two standard deviations. In our
previous paper, ref.~\citen{[1]} we concluded that the precise measurements 
of $\sigma_{\rm tot}^{pp}$ in the coming LHC measurements will resolve this 
discrepancy at $\sqrt s =1.8$TeV. 
It would still be worthwhile , however, to fix this problem in the
CDF and D0 experiments, since these values play an important role to search
for $\sigma_{\rm tot}^{(+)}$ and $\rho^{(+)}$ in the higher energy regions.

%\section*{Acknowledgements}
%We would like to thank ...........

\appendix
\section{Reanalysis of our predictions at the LHC ($\sqrt s$=14TeV) 
with $F^{(+)}(0)$ parameter} %Empty argument \section{} yields `Appendix'. 

In our previous work\cite{[1]}, 
we exploited the experimental data $\sigma_{\rm tot}^{(+)}$ 
and $\rho^{(+)}$ above $P_{lab}$=70GeV up to Tevatron energy ($\sqrt s=1.8$TeV)
to predict $\sigma_{\rm tot}^{(+)}$ 
and $\rho^{(+)}$ in the LHC region, based on Eq.~(\ref{eq8}) of $\rho^{(+)}$, not by Eq.~(\ref{eq9}). 
Although
the effect of the parameter $F^{(+)}(0)$ in the new formula (Eq.~(\ref{eq9}))
is not large 
in the high energy region, we show the results of the analyses based on Eq.~(\ref{eq9}) 
here for completeness.

\begin{figure}
\includegraphics{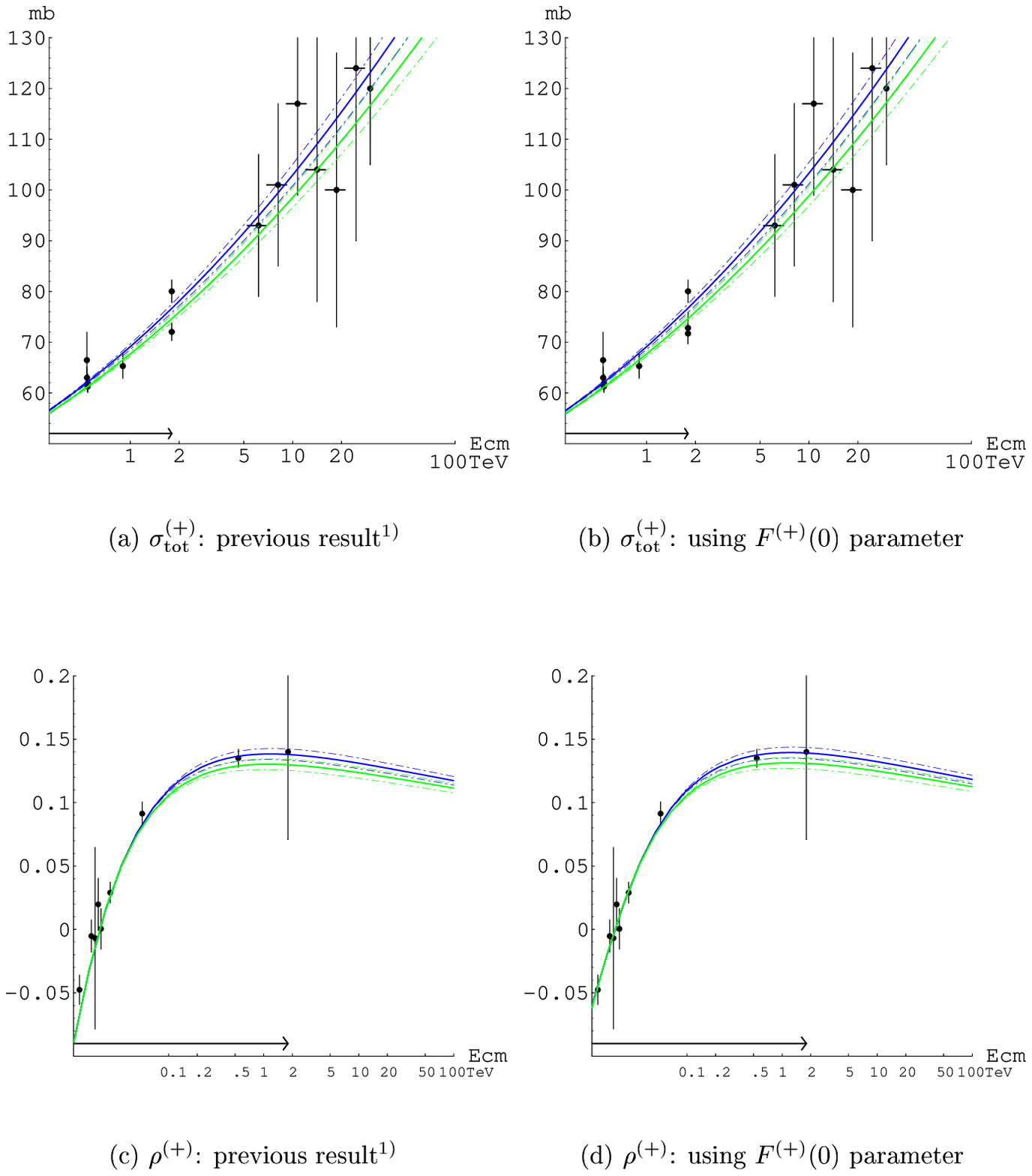}
\caption{\label{Ap} Predictions for $\sigma^{(+)}$ and $\rho^{(+)}$
compared with the previous results:
The new results using $F^{(+)}(0)$ parameter are shown by right figures, (b) and (d), respectively,
which are compared with the left figures, (a) and (c), of the previous analyses.\cite{[1]}
Predictions in terms of the fit 2(3) are shown by  
green(blue) lines, and  
the thin dot-dashed lines represent the one standard deviation of $c_2$.
(See the caption in Table \ref{tab1}.) 
The corresponding values of parameters are given in Table \ref{ap1}.  }
\end{figure}

Corresponding to ref.~\citen{[1]}
two independent analyses are done: one includes the E710$/$E811 data at $\sqrt s$=1.8TeV
denoted as fit 2 in ref.~\citen{[1]}, 
and the other includes the CDF datum of $\sigma_{\rm tot}^{(+)}$ at the same energy
denoted as fit 3 in ref.~\citen{[1]}. 
The results of the simultaneous fit to $\sigma_{\rm tot}^{(+)}$ 
and $\rho^{(+)}$ are compared with the previous results\cite{[1]}
in Fig.~\ref{Ap}. The fit to $\rho^{(+)}$ is slightly improved in the lower energy region, while 
the result of $\sigma_{\rm tot}^{(+)}$ is almost the same as the previous one.
The obtained values of parameters and the resulting $\chi^2$ are given in 
Table \ref{ap1} and Table \ref{ap2},
respectively.
\begin{table}
\caption{
The best-fit values of parameters in the 
fit 2 (fit up to Tevatron-collider energy including E710/811 data) and 
fit 3 (including CDF datum). 
The errors here correspond to the one standard deviation of $c_2$.
(See the caption in Table \ref{tab1} .) 
}
\begin{center}
\begin{tabular}{c|ccccc|}
       & $c_2$  & $c_1$  & $c_0$  &  $\beta_{P^\prime}$ & $F^{(+)}(0)$ \\
\hline
fit 2  & $0.0424\pm 0.0041$ & $-0.099\mp 0.069$ & $6.04\pm 0.28$ & $7.61\mp 1.55$ & $12.48\pm 0.73$\\
fit 3  & $0.0496\pm 0.0043$ & $-0.205\mp 0.072$ & $6.44\pm 0.29$ & $7.20\mp 0.81$ & $12.78\pm 0.72$\\
\hline
\end{tabular}
\end{center}
\label{ap1}
\end{table}

\begin{table}
\caption{ The values of $\chi^2$ for the fit 2 and fit 3. 
$N_F$ and $N_\sigma (N_\rho )$ are the degree of freedom and 
the number of $\sigma^{(+)}_{\rm tot}(\rho^{(+)})$ data points in the fitted energy region.
}
\begin{center}
\begin{tabular}{c|ccc|}
         & $\chi^2/N_F$        
            &  $\chi^2_{\sigma}/N_{\sigma}$ & $\chi^2_{\rho}/N_{\rho}$   \\
\hline
fit 2  & 11.6/22 & 7.9/18 & 3.7/9\\
fit 3  & 10.9/22 & 8.7/18 & 2.1/9\\
\hline
\end{tabular}
\end{center}
\label{ap2}
\end{table}

The fit to $\rho^{(+)}$ in the lower energy region is improved in comparison with the previous result,
as can be seen in Fig.~\ref{Ap}. Correspondingly 
much smaller $\chi^2_\rho$ is obtained in Table \ref{ap2}, which is 
compared with the previous values, $\chi^2_\rho$=8.4(6.9) for fit 2(3)\cite{[1]}. 

Predicted values of $\sigma_{\rm tot}^{(+)}$ and $\rho^{(+)}$
at LHC energy($\sqrt s$=14TeV) and at cosmic-ray energy ($P_{lab}$=$5\times 10^{20}$eV)
are given in Table \ref{ap3}.

\begin{table}
\caption{
The predictions of $\sigma_{\rm tot}^{(+)}$ and $\rho^{(+)}$ 
at the LHC energy $\sqrt{s}=E_{cm}=14$TeV($P_{lab}$=1.04$\times 10^8$GeV), and 
at a very high energy $P_{lab}=5\cdot 10^{20}$eV
($\sqrt s$=$E_{cm}$=967TeV.) 
in the cosmic-ray region. The errors correspond to one standard deviation of $c_2$.
}
\begin{tabular}{c|cc|cc|}
        &  $\sigma_{\rm tot}^{(+)}$({\scriptsize $\sqrt s$=14TeV}) 
        &  $\rho^{(+)}$({\scriptsize $\sqrt s$=14TeV})
        &  $\sigma_{\rm tot}^{(+)}$({\scriptsize $P_{lab}$=$5\cdot 10^{20}$eV}) 
        &  $\rho^{(+)}$({\scriptsize $P_{lab}$=$5\cdot 10^{20}$eV})\\
\hline
fit 2   & $104.2\pm 2.3$mb  & $0.123\pm 0.004$
        & $191\pm 8$mb  & $0.100\pm 0.003$\\
fit 3   & $109.3\pm 2.4$mb  & $0.130\pm 0.004$
        & $206\pm 8$mb  & $0.105\pm 0.003$\\
\hline
%our pred.   & $106.3\pm 5.1_{\rm syst} \pm 2.4_{\rm stat}$mb  
%               & $0.126\pm 0.007_{\rm syst}\pm 0.004_{\rm stat}$
%            & $196\pm 15_{\rm syst} \pm 8_{\rm stat}$mb 
%               & $0.102\pm 0.004_{\rm syst}\pm 0.003_{\rm stat}$\\
%\hline
\end{tabular}
\label{ap3}
\end{table}

The predictions combining the two results in Table \ref{ap3} are
\begin{eqnarray}
\sigma_{\rm tot}^{\bar pp} &=&  106.8\pm 5.1_{\rm syst} \pm 2.4_{\rm stat}\ {\rm mb},\ \   
\rho^{\bar pp} = 0.127\pm 0.007_{\rm syst}\pm 0.004_{\rm stat}\ \ \ \ \ \ \ \nonumber\\ 
\sigma_{\rm tot}^{pp} &=& 198\pm 16_{\rm syst} \pm 8_{\rm stat}\ {\rm mb},\ \    
\rho^{pp} = 0.103\pm 0.004_{\rm syst}\pm 0.003_{\rm stat}\ \ \ \ \ \ \ 
\label{Aeq1}
\end{eqnarray}
at the LHC energy($\sqrt s=E_{cm}=14$TeV) and the cosmic-ray energy ($P_{lab}=5\times 10^{20}$eV), 
respectively. 
The above results are almost the same as the previous ones, 
Eq.~(13) of ref.~\citen{[1]}. 
%Systematic errors are taken as differences of two predictions in the best fits.
Here we obtain fairly large systematic uncertainty again
coming from the data treatment at the Tevatron-energy.

%\section{Second Appendix}

\end{document}